\documentclass{moriond}

\usepackage{amsmath}
\usepackage{graphicx}
\usepackage{multirow}
\usepackage{ulem} %added to be able to cross text out
\usepackage{units} % for using nicefrac in texts
\usepackage{version} % for comment
\usepackage{color}
\usepackage{colortbl}
\usepackage{xcolor}  % for writing parts in color, e.g. in red
\usepackage{url} 

\def\be{\begin{equation}}
\def\ee{\end{equation}}
\def\bea{\begin{eqnarray}}
\def\eea{\end{eqnarray}}

\newcommand{\ms}{\ \text{ms}}
\newcommand{\m}{\ \text{m}}
\newcommand{\s}{\ \text{s}}

\newcommand{\Hz}{\ \text{Hz}}

\newcommand{\Photo}{}

\begin{document}
\vspace*{4cm}
\title{Matter-wave laser Interferometric Gravitation Antenna (MIGA): \\ 
New perspectives for fundamental physics and geosciences}

\author{R. Geiger$^1$, L. Amand$^1$, A. Bertoldi$^2$, B. Canuel$^2$, W. Chaibi$^4$, C. Danquigny$^{5,6}$, I. Dutta$^1$, B. Fang$^1$, S. Gaffet$^3$, J. Gillot$^2$, D. Holleville$^1$, A. Landragin$^1$, M. Merzougui$^4$,  I. Riou$^2$, D. Savoie$^1$ and P. Bouyer$^2$}

\address{$^1$SYRTE, Observatoire de Paris, PSL Research University, CNRS, Sorbonne Universit\'es, UPMC Univ. Paris 06, LNE, 61 avenue de l'Observatoire, 75014 Paris, France}

\address{$^2$LP2N, Laboratoire de Photonique Num\'erique et Nanosciences, Institut d'Optique Graduate School IOA, Rue Fran{\c c}ois Mitterrand, 33400 Talence, France}

\address{$^3$LSBB, UMS UNS, UAPV, CNRS, 84400 Rustrel, France}

\address{$^4$ARTEMIS - Observatoire de la C{\^o}te d'Azur, Boulevard de l'Observatoire CS 34229, 06304 Nice Cedex 04, France}

\address{$^5$UAPV, UMR1114 EMMAH, F-84000 Avignon, France}

\address{$^6$INRA, UMR1114 EMMAH, F-84914 Avignon, France}

\maketitle
\abstracts{
The MIGA project aims at demonstrating precision measurements of gravity with cold atom sensors in a large scale instrument and at studying the associated applications in geosciences and fundamental physics. The first stage of the project (2013-2018) will consist in building a 300-meter long optical cavity to interrogate atom interferometers and will be based at the low noise underground laboratory LSBB in Rustrel, France. The second stage of the project (2018-2023) will be dedicated to science runs and  data analyses in order to probe the spatio-temporal structure of the local gravity field of the LSBB region, a site of high hydrological interest. MIGA will also assess future potential applications of atom interferometry to gravitational wave detection in the frequency band  $\sim 0.1-10$ Hz hardly covered by future long baseline optical interferometers.
This paper presents the main objectives of the project, the status of the construction of the instrument and the motivation for the applications of MIGA in geosciences. Important results on new atom interferometry techniques developed at SYRTE in the context of MIGA and paving the way to precision gravity measurements are also reported. 
}

\section{Introduction}

After more than 20 years of fundamental research, atom interferometers (AIs) have reached sensitivity and accuracy levels competing with or beating inertial sensors based on different technologies. Atom interferometry offers interesting applications in geophysics (gravimetry, gradiometry, Earth rotation rate measurements), inertial sensing (submarine or aircraft autonomous positioning), metrology (new definition of the kilogram) and fundamental physics (tests of the standard model, tests of general relativity). AIs already contributed significantly to fundamental physics by, for example, providing stringent constraints on quantum-electrodynamics through measurements of the hyperfine structure constant~\cite{Bouchendira2011}, testing the Equivalence Principle with cold atoms~\cite{Schlippert2014}, or providing new measurements for the Newtonian gravitational constant~\cite{Rosi2014}. Cold atom sensors have been established as key instruments in metrology for the new definition of the kilogram \cite{Thomas2015} or through international comparisons of gravimeters~\cite{Gillot2014}. 
The field of atom interferometry is now entering a new phase where very high sensitivity levels must be demonstrated, in order to enlarge the potential applications outside atomic physics laboratories. These applications range from gravitational wave (GW) detection in the $[0.1-10 \Hz]$ frequency band to next generation ground and space-based Earth gravity field studies to precision gyroscopes and accelerometers. 

The Matter-wave laser Interferometric Gravitation Antenna (MIGA) project will explore the use of atom interferometry techniques to build a large-scale matter-wave sensor which will open new applications in geoscience and fundamental physics.
The MIGA consortium gathers 15 expert French laboratories and companies in atomic physics, metrology, optics, geosciences and gravitational physics, with the aim to build a large-scale underground atom-interferometer instrument by 2018 and operate it till at least 2023. 
In this paper, we present the main objectives of the project, the status of the construction of the instrument and the motivation for the applications of MIGA in geosciences. Important results on new atom interferometry techniques developed at SYRTE in the context of MIGA and paving the way to precision gravity measurements are also reported.

\section{MIGA principle and sensitivity}
\label{sec:principle_sensitivity}

\begin{figure}[h!]
\centering
\includegraphics[width=\linewidth]{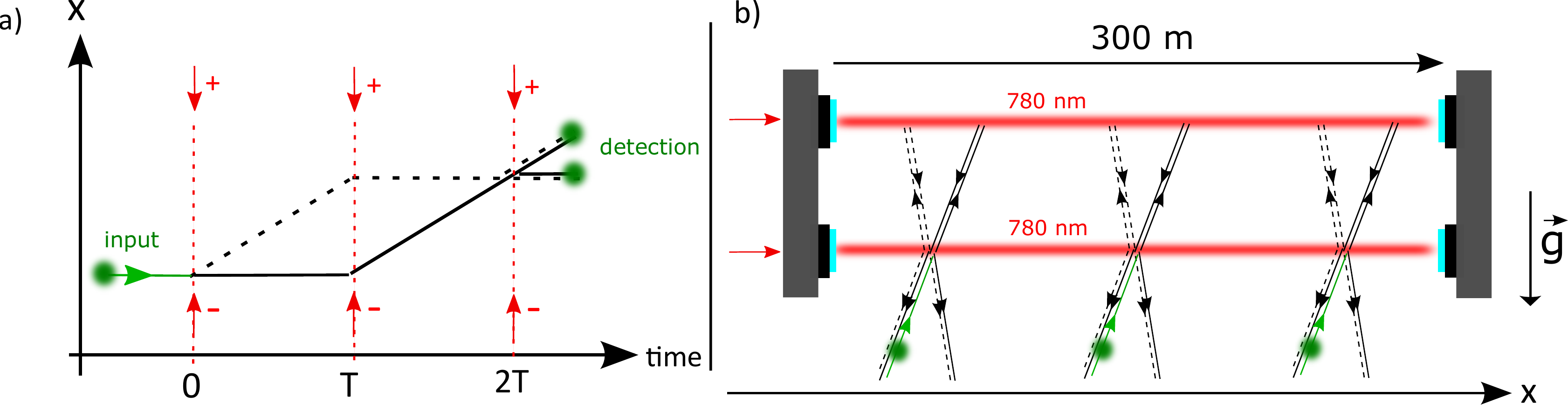}
\caption{\textbf{MIGA geometry.} a) Schematic of the 3 pulse AI. Two laser beams propagating in opposite directions are used to split and recombine the matter waves. The atoms are detected at the output using fluorescence detection. b) Sketch of the MIGA baseline: 3 AIs are interrogated by the 780 nm laser beams resonating inside two two horizontal optical cavities, placed in a vertical plane and with a distance set by the interrogation time.}
\label{fig:AI_geometry}
\end{figure}

The AI geometry of MIGA is similar to the one of a Mach-Zehnder Interferometer for optical waves. The geometry is described in
Fig.~\ref{fig:AI_geometry}a) where matter waves are manipulated by a set of counter-propagating laser pulses. At the input of the interferometer, a
$\pi/2$ pulse creates an equiprobable coherent superposition of two different momentum states of the atom. The matter-waves are then deflected by the
use of a $\pi$ pulse before being recombined with a second $\pi/2$ pulse. To realize these beam-splitters and mirror pulses, MIGA will make use of Bragg
diffraction of matter-waves on light standing waves \cite{Martin1998}. Conservation of energy-momentum during this process imposes to couple only atomic
states of momentum $|+\hbar k\rangle$ to state $|-\hbar k\rangle$ where $k=2\pi\nu_0/c$ is the wave vector of the interrogation field. At the output of
the interferometer, the transition probability between these states is obtained by a two waves interference formula $P = \frac{1}{2}(1 + \cos\phi)$. The
atom phase shift $\phi$ depends on the  phase difference between the two couterpropagating lasers which is imprinted on the diffracted matter-wave
during the light pulse. MIGA will make use of a set of such AIs interrogated by the resonant field of an optical cavity as described in
Fig.~\ref{fig:AI_geometry}b). In this configuration, each AI will measure the inertial effects $s_X(X_i)$ along the cavity axis at position $X_i$
together with GW effects associated to the cavity propagation of the interrogation laser. Spurious effects such as fluctuations of the cavity mirror
position $x_1(t)$ and $x_2(t)$ or laser frequency noise $\delta\nu(t)$ also affect the AI signal. Taking into account these different effects, the atom
phase shift $\phi(X_i)$ measured by the AI at position $X_i$ reads:
\be
\phi(X_i) = 2k s_{x2} +2k \left(\frac{s_{\delta\nu}}{\nu_0} + \frac{s_h}{2}\right)(X_i-L) + 2k s_X(X_i)
\label{eq:AIsignal}
\ee
where $h$ is the GW strain amplitude, $L$ is the mean cavity length and $s_u$ accounts for the convolution of the time-fluctuations of effect $u(t)$ by the AI sensitivity function $s(t)$ \cite{Cheinet2008IEEE}. The last term in Eq.~\eqref{eq:AIsignal} accounts for the acceleration of the  center of mass of the free falling atom cloud which  depends on the local position of the AI because of the non-homogeneous gravitational field.
Common mode rejection between the AI signals at different positions will enable to cancel
out most of the contribution of cavity mirror position fluctuations $s_{x2}$. The influence of laser frequency noise
will be kept negligible in the first version of the MIGA instrument (till 2023)  by using state-of-the-art ultra stable laser techniques, yielding relative stabilities better than $\delta\nu/\nu_0\sim 10^{-15}$. 
Eq.~\eqref{eq:AIsignal} thus shows that the instrument can be used for local monitoring of mass motion encoded in the last term $s_X$ and, in the future, for GW detection without being affected by position noise of the optics.

The last term of Eq.~\eqref{eq:AIsignal} can be written as $\phi(X)=2k a(X) T^2$ where $a(X)$ is the local acceleration of gravity of the atoms at position $X$ and $T$ is the time between the light pulses in a 3 light pulse geometry (see Fig.~\ref{fig:AI_geometry}a)). For two AIs separated by the baseline $L=X_2-X_1$ and assuming a constant gravity gradient $\gamma$, the gradiometer sensitivity of the instrument is given by  $\sigma_\gamma=\sigma_\phi/(2kLT^2 \sqrt{\tau})$ where $\sigma_\phi$ is the AI phase sensitivity and $\tau$ the integration time. For $L=100 \m$, a shot-noise limited AI with $10^6$ atoms (1 mrad phase sensitivity) and $T = 0.5 \ \text{s}$ yields a gradiometer sensitivity of $2.4\times 10^{-13} \ \text{s}^{-2}$ after $\tau=100 \s$ of integration time, sufficient to detect a mass anomaly of few tons in a 100 meter region around the instrument. Combining the different measurements provided by the array of AIs will bring to a more precise measurement of the mass anomaly position and potentially movement.

\section{MIGA subsystems}

\subsection{Cold atom source}

\begin{figure}[h!]
\centering
\includegraphics[width=0.7\linewidth]{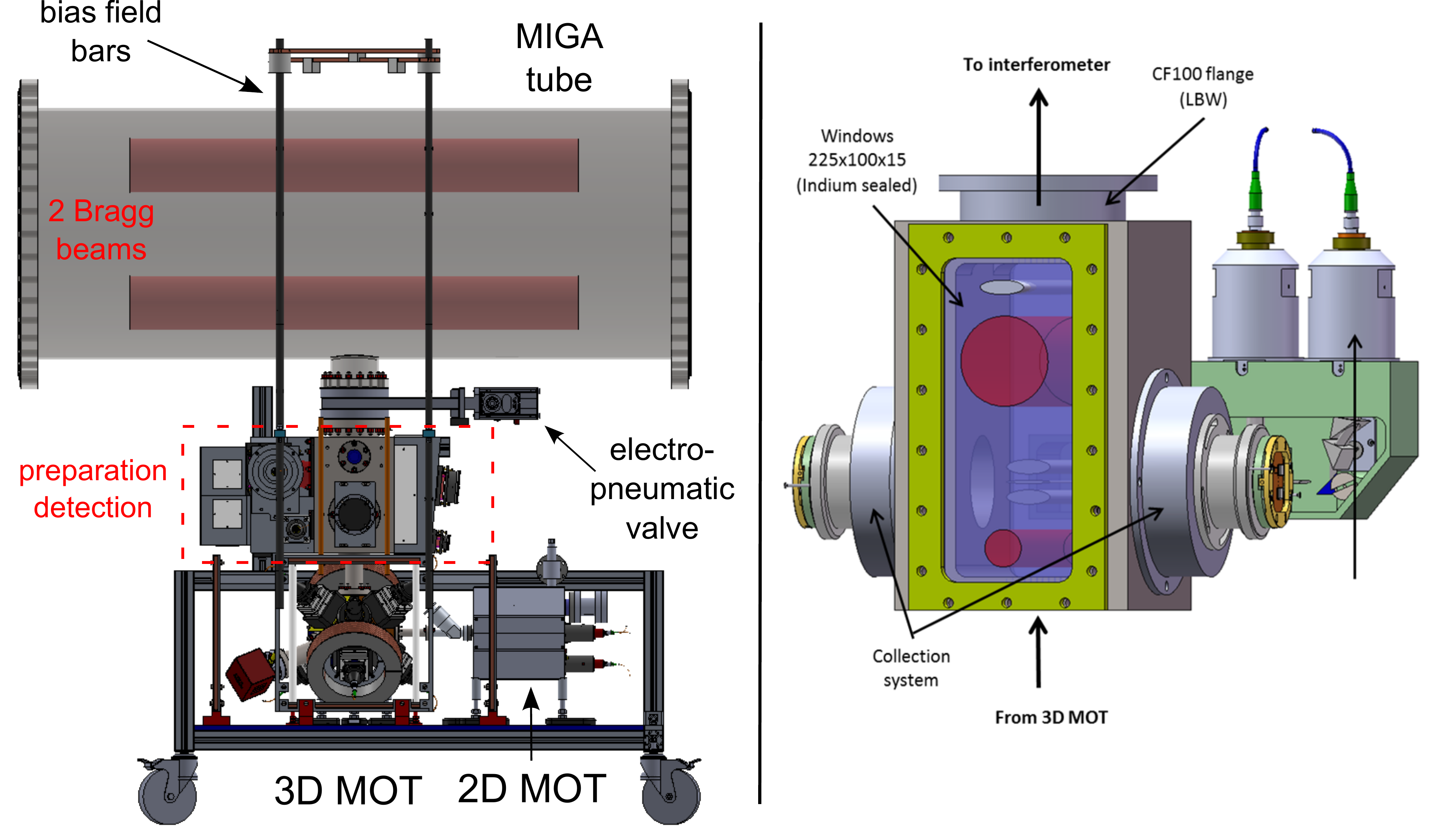}
\caption{Left: Global view of the cold atom source unit. Right: technical drawing of the cold atom preparation and detection region.
}
\label{fig:cold_atom_source}
\end{figure}

The atom source unit delivers cold atom clouds which will be interrogated by the MIGA cavity Bragg beams to form the AI. The general design of the unit is presented in Fig.~\ref{fig:cold_atom_source}. Its main functions are \textit{(i)} the loading and laser cooling of $^{87} \text{Rb}$ atoms, \textit{(ii)} the launching of the atomic cloud along a quasi-vertical trajectory and the control of the angle of the trajectory with respect to the cavity beams, \textit{(iii)} the preparation of the cold atom source before it enters the interferometer, and \textit{(iv)} the detection of the atoms at the interferometer output. 

In order to optimize the contrast of the AI, the quantum state of the atoms is prepared on their way up, before the interrogation region. 
A first counter-propagating velocity-selective Raman pulse (bottom red beam in Fig.~\ref{fig:cold_atom_source}) is used to select the atoms in the $m_F=0$ Zeeman sub-level of the $F=2$ hyperfine state, with a relatively narrow velocity class (width of 1 photon recoil, corresponding to a temperature of $\approx 400$ nK in the direction of the Raman lasers).
The unselected atoms are then pushed by a laser tuned on resonance with the cycling transition. 
This Raman/push procedure  is repeated a second time to clean the remaining unwanted atoms produced by spontaneous emission on the first Raman selection pulse. For this purpose, we use the Raman 2 beam (top big red beam) with approximately the same duration and Rabi frequency as the Raman 1 beam to transfer the atoms back to the $F=2$ state. The remaining atoms in the $F=1$ state are pushed with an orthogonal push beam tuned on the $F=1\rightarrow F^\prime= 0$ transition (gray beam at the top).
The angle of the Raman beams can be tuned by few degrees around zero in order to introduce a Doppler effect which allows lifting the degeneracy between the $|p\rangle\rightarrow |p+2\hbar k\rangle$ and $|p\rangle\rightarrow |p-2\hbar k\rangle$ transitions. In this way, the atoms will enter the interferometer in a well-defined momentum state. Moreover, the Raman beam angle enables to control the angle of the trajectory with respect to the vertical direction, i.e. the Bragg angle.
After this all-optical preparation steps, the atoms  enter the interferometer in the $|F=2,m_F=0\rangle$ internal state, with a relatively narrow velocity distribution in the longitudinal direction of the Bragg interrogation beams and with a well-controlled trajectory.

After their interrogation by the Bragg beams in the AI, the two different momentum states $|\pm\hbar k\rangle$ of the atoms are
labelled to two different internal states with the Raman 2 laser. More precisely, the velocity selective feature of the Raman transition is used to
transfer the $|F=2,\hbar k\rangle$ atoms to the $F=1$ internal state, while the $|F=2,-\hbar k\rangle$ atoms remain in the $F=2$ internal state. The
atoms can then be resolved with common fluorescence techniques.  Detection of the atoms labelled in $F=2$ is first realized with a light sheet beam (see
Fig.~\ref{fig:cold_atom_source}) tuned on resonance on the $F = 2 \rightarrow F^\prime= 3$ transition. The beam is partially blocked at the
retroreflection mirror so that the atoms acquire a net momentum in the beam direction and will therefore not be resonant with the following light beams.
The $F = 1$ atoms are re-pumped to the $F = 2$ state using a thinner intermediate light sheet, before these $F=2$ atoms enter the third light sheet. The
fluorescence light of the two light sheets is collected by a $2 \%$ collection efficiency lens and imaged on a photodiode. The fluorescence signal is
used to reconstruct the normalized atomic populations and then the transition probability, yielding the AI phase.

\subsection{Laser System.}
The different lasers used to cool and manipulate the atoms are delivered from an all-fibered laser module developed by the company $\mu$Quans \cite{muquanswebsite}. The laser architecture is based on frequency doubled telecom lasers, as already described in various publications, see e.g. Ref.~\cite{Menoret11}. A Master laser is locked using a  Rubidium 85 saturated absorption spectroscopy signal and references 3 slave diodes which are respectively used for the 2D MOT cooling laser, the 3D MOT cooling/Raman 2 laser, and the 3D MOT repumper/Raman 1 laser. The 3 slave diodes are all phase locked to the Master laser. The repumping light for the 2D MOT is generated by a fiber electro-optic phase modulator at 1560 nm fed with the appropriate microwave frequency.

After amplification in Erbium doped fiber amplifiers and second harmonic generation in PPLN waveguide crystals, the 780 nm light is sent to optical splitters and guided to the experiment chamber using several optical fibers. The laser module nominally delivers 170 mW total power for the 2D MOT, 150 mW total power for the 3D MOT, and $175 \ \text{mW}$ in each of the two Raman beam couple, used for the preparation stage and the detection respectively. The  power and polarization fluctuations at the fiber outputs are close to the one percent level.
The phase lock signals are controlled by radio and microwave frequency sources all referenced to a stable 100 MHz oscillator.
The full laser system is hosted in a $1.7\times 0.5\times 0.5 \ \text{m}^3$ transportable rack.

\subsection{Optical cavity setup}

\begin{figure}[h!]
\centering
\includegraphics[width=\linewidth]{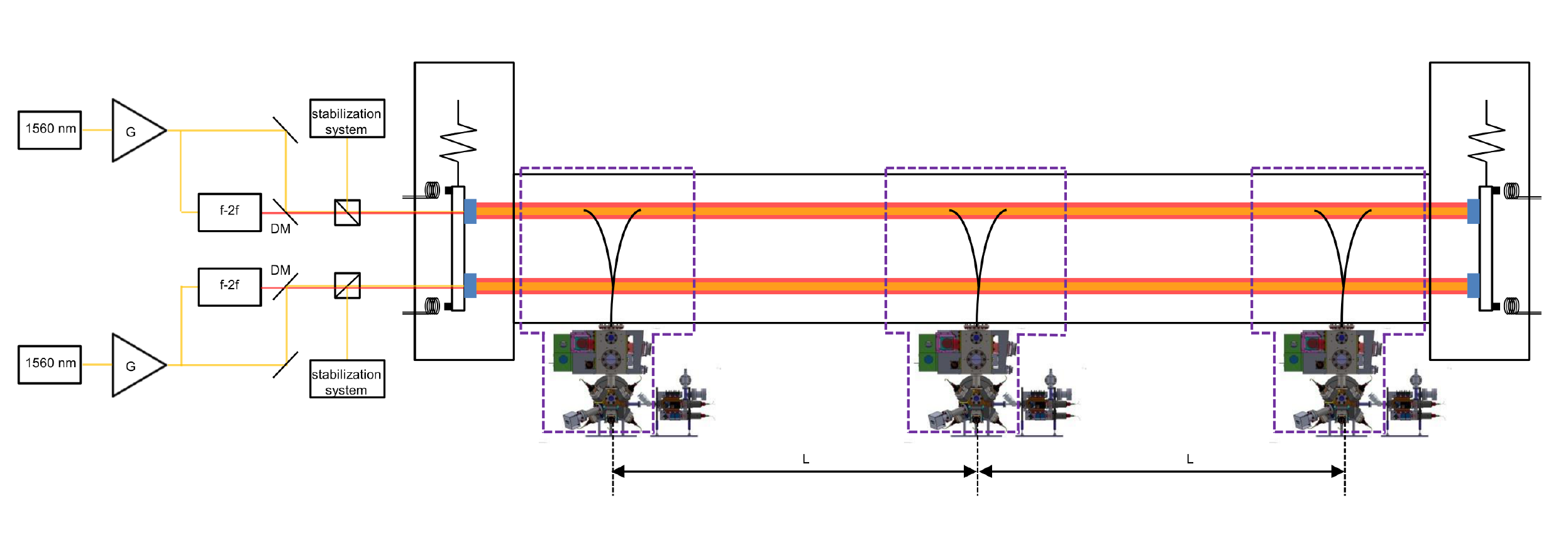}
\caption{\textbf{Overview of the MIGA cavity with the main sub-systems.} The three
atomic heads separated by a distance $L$ launch atomic clouds in an almost vertical
parabolic flight. The atoms are manipulated in the upper part of the
parabola with a Bragg interferometric sequence by way of radiation pulses at 780
nm (red lines) resonant with two horizontal cavities. The resonance condition for the
interrogation light relies on generating the 780 nm via frequency doubling of a 1560
nm laser (yellow lines) locked to one of the two cavities, and using stabilized, common
payloads for the mirrors on each side of the cavities to avoid relative length fluctuations
of the two resonators. The Ultra High Vacuum system encompassing the optical cavities, the mirror
payloads and their stabilization system is represented with gray solid lines; to it are
connected the atomic source units. Each interferometric region and most of the related
atomic head are enclosed in a $\mu$-metal shield, represented with dashed violet lines. The
control system of the experimental setup and the laser systems dedicated to each
atomic head are not represented in the plot.
}
\label{fig:global_setup}
\end{figure}

The intensity of the Bragg pulses is enhanced thanks to two cavities, one for the
splitting and projection $\pi$/2 pulses, and one at the trajectories' apogee for the $\pi$ pulse.
The solution adopted to have the 780 nm interrogation pulses at resonance with the
cavities relies on obtaining the probe radiation via frequency doubling of a telecom
laser at 1560 nm continuously locked to one of the two resonators to track its length
variations. The servo system is used also to control the payload tilts and rotations so
as to maintain the phase coherence between the two cavities. The Bragg pulses are
shaped with acousto-optic modulators (AOMs) on the two beams at 780 nm before
their injection in the cavities. The telecom laser is phase modulated and locked to
the cavity on one frequency sideband, and the modulation frequency $\Omega$ is chosen so as to have
the doubled component of the carrier resonant with the resonator. 
$\Omega$ has to account for
the different cavity length at 780 nm and 1560 nm, because of the refraction index of
the two coatings on the mirrors, as well as for the frequency shift imposed by the AOM
used to pulse the interrogation beams. The LP2N laboratory is currently developing a prototype system
using low power laser sources (100 mW at 1560 nm, and 1 W at 780 nm). The  CELIA laboratory at Bordeaux 1 University is
developing a high power solution, which targets 100 W of radiation at 780 nm before the
injection in the cavity.

The two cavities share a common payload on each side to hold the mirrors, placed
at a vertical distance of $\approx 30.6$ cm to have an interrogation time $T=250 \ \text{ms}$. The
impact of ground seismic noise on the position of the cavity mirrors will be reduced by
way of an anti-vibration system, which must limit the related phase noise contribution
on each AI. Two different approaches are being considered: a passive system of mechanical filters to suspend each payload, and
an active stabilization of each mirror position using piezoelectric actuators. The main
constraints on the system are set by the level of the seismic noise at the installation site
(LSBB), and the response function of the AIs to mirror acceleration noise, as in Eq.~\eqref{eq:AIsignal}.

\section{High sensitivity atom interferometry techniques}

The performance of the MIGA antenna will rely on the possibility to achieve high sensitivity gravito-inertial measurements. Moreover, future applications to gravitational wave detection will require higher bandwidth ($\sim 10 \Hz$) cold AIs than what is currently obtained in laboratories (about $1 \Hz$ for long $T$ AIs). In this section, we briefly present two results obtained at the SYRTE laboratory on an AI which geometry is similar to the fountain-like architecture of the MIGA sensors. 

\begin{figure}[h!]
\centering
\includegraphics[width=\linewidth]{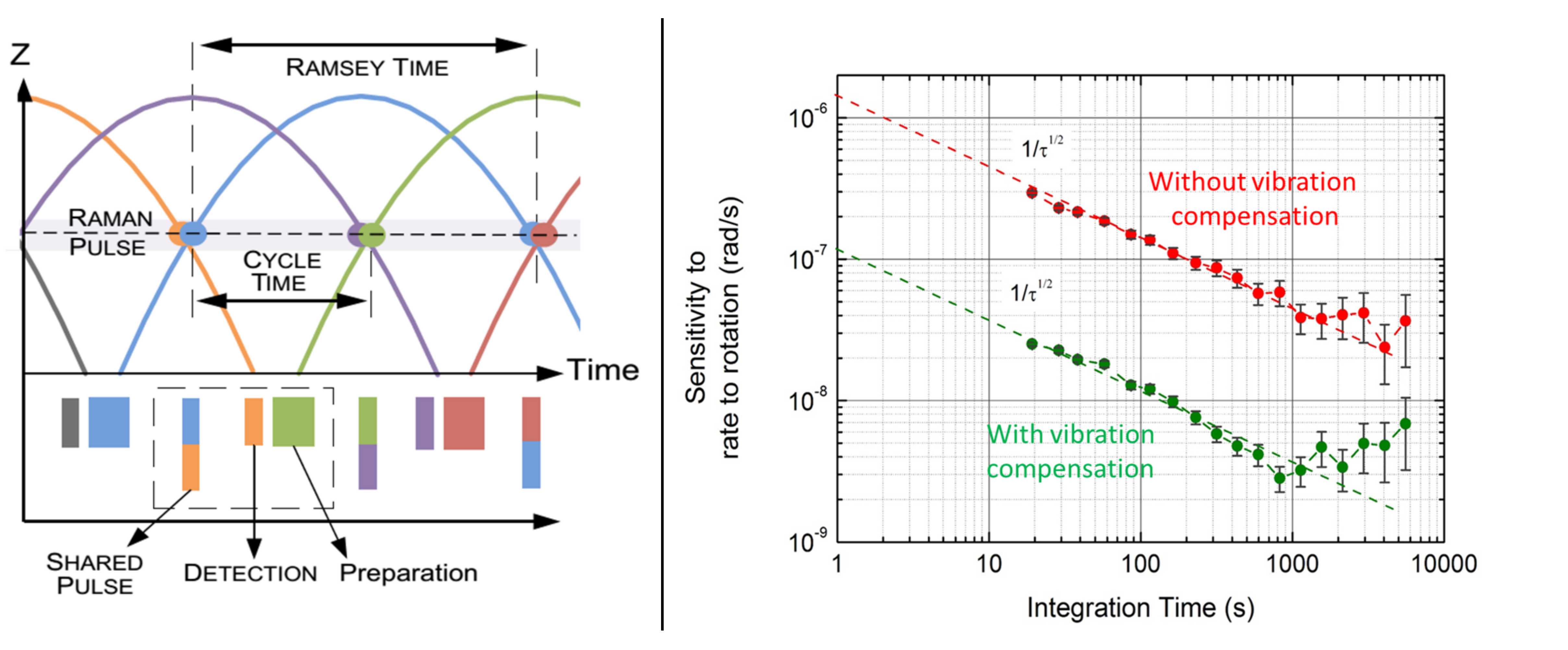}
\caption[]{\textbf{New atom interferometry techniques for high precision inertial measurements.} Left: schematic of the joint interrogation technique allowing the interrogation of several clouds of atoms simultaneously in the interferometer and rejection of vibration noise aliasing due to dead times. From Ref.~\cite{Meunier2014}. Right: SYRTE cold atom gyroscope with a $3 \ \text{nrad/s}$ after $1000 \s$ of integration time using a Sagnac matter-wave interferometer of $2.4$ cm$^2$ area \cite{DuttaInPrep,Barrett2014}.}
\label{fig:AI_techniques}
\end{figure}

First, we demonstrated a new method to interrogate several clouds of cold atoms simultaneously in the interferometer in a so-called joint interrogation scheme~\cite{Meunier2014} (the principle of the joint interrogation is represented in Fig.~\ref{fig:AI_techniques}, left panel). 
Conventional cold AIs run in sequential mode: after laser cooling, the cold atoms are injected in the interferometer where the inertial effects are measured. Thus, the sensor does not operate continuously.  Information on signals varying during the cold atom source preparation is lost, which is a major drawback for various applications. Moreover, the aliasing effect of the vibration noise associated with the dead time results in a degradation of the short term sensitivity.
To circumvent this problem, the joint interrogation solution compatible with the MIGA fountain geometry  allows interrogating the atoms in the interferometer region, while another cold atom cloud is being prepared. This leads to a zero-dead time gyroscope. Moreover, if the different atom clouds share common (Bragg) interrogation pulses, the vibration noise is correlated between the successive measurements, which leads to a faster averaging of the vibration noise.
 We demonstrated a multiple joint operation in which up to five clouds of atoms were interrogated simultaneously in a single fountain with $2T = 800 \ms$ interrogation time \cite{Meunier2014}. The essential feature of the multiple joint operation, which we demonstrated for a micro-wave Ramsey interrogation is currently being generalized to the inertial sensor operation. The multiple joint operation gives access to high-frequency components while maintaining high sensitivity linked to long interaction times achievable with cold atom sensors.

A second key feature of MIGA will be to operate an AI with a long interrogation time, $2T= 500 \ms$, yielding a high accelerometer scale factor $2k T^2$. In this regime, the effect of vibrations from the payload results in several radians of AI phase noise and must be managed to keep it below the targeted phase sensitivity level (ideally below the atom shot noise of $\sim 1 \ \text{mrad}$). In the MIGA gradiometer configuration the vibration noise is partially common to the different AIs, so the differential phase can be obtained by adopting specific retrieval methods~\cite{Sorrentino2012,Barrett2015}. Nevertheless, the extraction of the interferometric phase still requires the AI to be operated in its linear range, i.e. $[0-\pi]$. Moreover, being able to extract the individual AI phase yields the absolute local gravity field in the direction of the Bragg beams and therefore provides additional information to the gravity gradient and its curvature. 
To this end, we demonstrated in the SYRTE experiment the possibility to reject the vibration noise with a factor up to 20 using classical accelerometers in an interferometer with $2T=800 \ms$  interrogation time. This noise rejection was performed in a gyroscope configuration where the AI mainly senses rotation rates, allowing us to demonstrate a gyroscope with $3 \ \text{nrad/s}$ long term stability (see Fig.~\ref{fig:AI_techniques}, right panel). These results strongly support the possibility to obtain high sensitivity gravity measurements with the MIGA interferometers.

\section{MIGA: new perspectives in geosciences}

\subsection{Hydrological interest of \textit{the Fontaine de Vaucluse/LSBB} site} 
\begin{figure}[h!]
\centering
\includegraphics[width=0.7\linewidth]{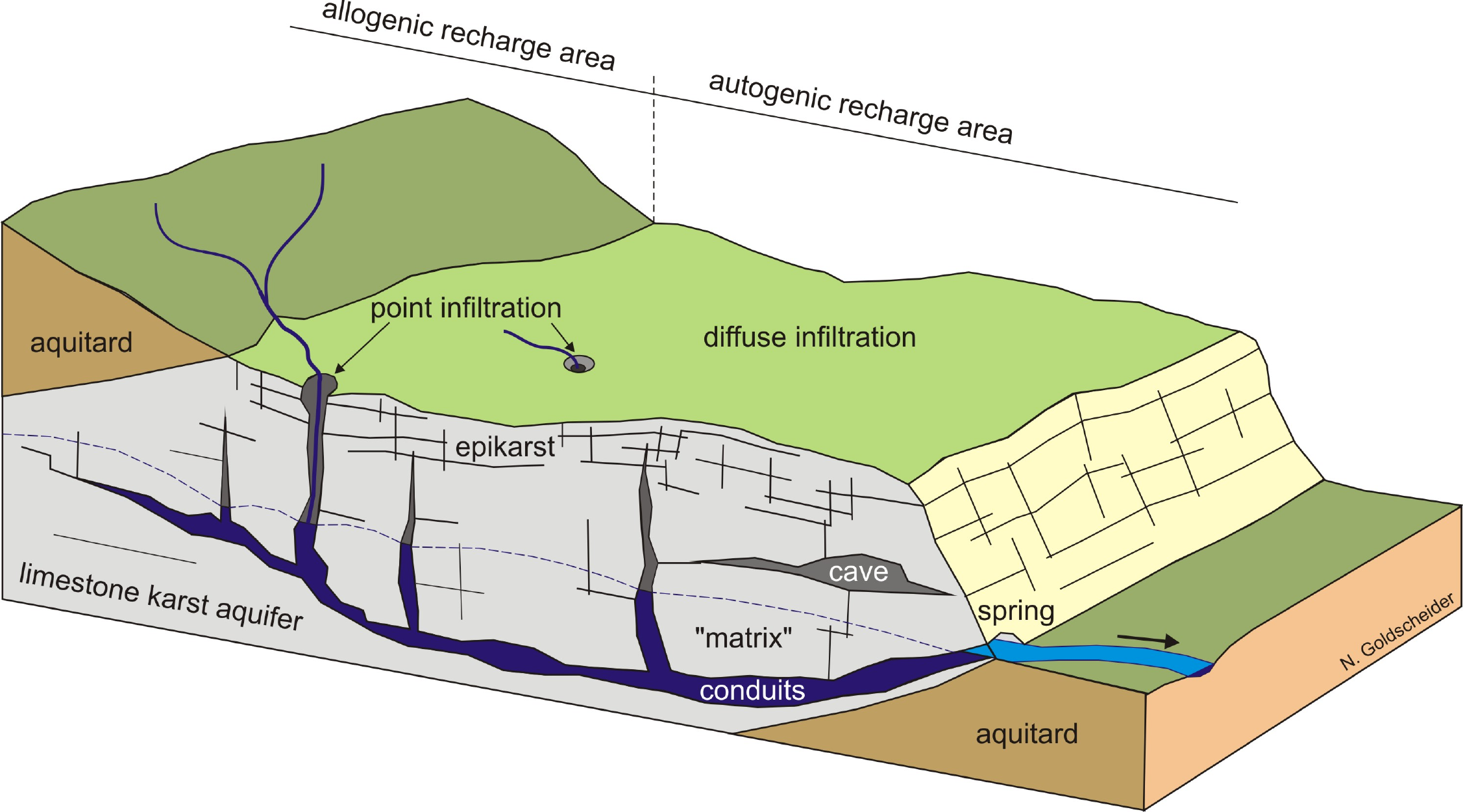}
\caption{Block diagram illustrating the hydro-geological functioning of a karst aquifer, from N Goldscheider, D Drew (2007), \textit{Methods in Karst Hydrogeology}, Taylor et Francis, \url{http://www.agw.kit.edu/english/3851.php}.
}
\label{fig:hydro}
\end{figure}

Almost a quarter of the world population obtains its drinking water
from karst hydrosystems~\cite{Ford2007} (see Fig.~\ref{fig:hydro} for a schematic). Efficient protection and sustainable management of such resources
require appropriate tools and strategies to be developed~\cite{Mudarra2010}. The numerical modelling of karst aquifers is probably the major stumbling
block in developing such tools. Karst remains aside from other hydrosystems, because the paroxysmal~\cite{deMarsily1984} and self-organized
\cite{Worthington2009} heterogeneity of that medium limits the relevance of classical hydro-geological tools, such as physically-based and gridded flow
models, and because of the difficulty of characterization of this heterogeneity. Hopefully, recent improvements of computing power and computational
techniques in the one hand and geophysical measurement techniques in the other hand~\cite{Berkowitz2002} enable considering now the applicability of
physically-based and gridded flow models to karst hydrodynamics. Developing and calibrating such tools requires the acquisition of 4D hydro-geo-physical
data (water content, flux and velocities, ...) at different scales.

In south-eastern France, the Fontaine-de-Vaucluse karst hydrosystem is one of the biggest karst watershed in the world: its catchment area is around  $1115$ km$^2$ and composed of a nearly 1500 m thick massive and continuous limestone~\cite{Masse1976} from Necomanian marls to upper Aptian marls. The Fontaine-de-Vaucluse spring is quite the only outlet of this hydrosystem and the biggest karst spring in Europe with an average outlet discharge of 19 m$^3$/s between 1877 and 2004~\cite{Cognard-Plancq2006}. Within this peculiar karst hydrosystem, LSBB (Low Noise Underground Research Laboratory) is an almost horizontal tunnel coming across the karst medium and intersecting arbitrarily faults, karst networks and flowpaths at depths between 0 and 519 m. All these elements make the Fontaine-de-Vaucluse and LSBB sites a relevant multi-scales observatory to develop physically-based and gridded flow models to karst hydrodynamics based on innovative 4D hydro-geo-physical data acquisition.

\subsection{Methods}

Whereas recent developments of geophysical methods enable to expect better characterization of complex hydrosystems~\cite{Berkowitz2002}, their application to karst remains not obvious~\cite{Chalikakis2011}. Nevertheless, various conventional techniques and instruments are currently applied to karst hydro-geology such as Electrical Resistivity Tomography (ERT) and 2D Ground Penetration Radar (GPR). One of the important questions is to have enough resolution and depth of investigation all at once to detail all the features controlling the groundwater circulation and storage from matrix porosity or micro-fracturing to major faults and karst conduits . On the other hand, estimating the variation of water mass requires the use of integrating methods directly or indirectly related to water content such as seismic, ERT, Magnetic Resonance Sounding (MRS) or Gravimetry. For instance, as shown in  Ref.~\cite{Carriere2013}, GPR results supply a near surface high resolution ($\sim 10 \ \text{cm}$) imaging and thus can provide relevant geological information such as stratifications and fractures. However, GPR's investigation depth remains limited to around 12 meters.
ERT surveys shows strong lateral and vertical variations which can inform on general geological structuring and feature orientation. ERT is able to prospect down to 40 meters but is a low resolution integrative technique.
Finally, active seismic reflection imaging or transmission tomography (at frequencies $\sim 10-500 \ \text{Hz}$) allows measuring the ground seismic velocities and probing the rock elasticity and porosity, yielding information on the rock structure and fractures~\cite{Maufroy2014}. The corresponding resolution is $\sim 10 \ \text{m}$ with accessible depth $\sim 100 \ \text{m}$.

In contrast to these techniques requiring an inversion model, atom interferometry can provide direct measurements of the surrounding mass distribution and thus represents an interesting complementary method. Moreover, long-term (years) measurements of the gravity field can be obtained thanks to the long term stability of cold atom sensors. In this context, MIGA will provide non-invasive measurements of the gravity field on a few 100 m scale. As discussed in section~\ref{sec:principle_sensitivity}, the typical gravity gradient sensitivity of MIGA will be $\sim 10^{-13} \ \text{s}^{-2}$ after 100 s of measurement time, with a maximum sensitivity in the direction of the baseline. Such gravity gradients correspond for example to a 1 ton mass at 100 meter from the instrument. If the source mass producing such gravity gradients moves in time, the AI antenna signal will vary accordingly. The spatial resolution of the antenna will depend on the number of AIs and their relative distance. Tuning the AI geometry (interrogation time $T$, number of pulses, etc.) allows changing the response of the sensor to the source mass and thus yielding more information.

\section{Conclusion and perspectives}

The MIGA instrument will use long baseline (300 m) optical and matter-wave interferometry for high precision gravity field measurements, in order to monitor subsurface mass transfers in the LSBB region, which represents a unique site of hydrological interest. Combining conventional instruments and methods from hydro-geology with cold atom gravitation sensor measurements will allow better modelling of karst aquifers, for which only very few (3+1)-dimensional data are currently available to constrain the models. MIGA will also investigate the applications of atom interferometry to extend the sensitivity of future GW detectors at frequencies below $10 \Hz$.

The first cold atom source unit has been characterized and will be installed at LP2N (Talence) in June 2015, where the first atom interferometry experiments in a 1 meter prototype optical cavity cavity will be performed. A 6 meter AI gradiometer prototype based at LP2N is currently under design and will allow  testing at a reduced scale the measurement strategy for mass monitoring at LSBB. Digging of the galleries at LSBB is planned for 2016, with an installation of the 300 meter long vacuum cavity and the three AI units in 2017. Following the final optimizations of the instrument, the operation phase should start in 2018.

The MIGA Equipex aims at being the first step to a larger, more ambitious program that may lead to a future European infrastructure. The development,
the scientific operation and the technical implementation of this first version of the gravitational antenna will pave the way to a more sensitive
version that will take advantage of the current fundamental research in atom interferometry. The MIGA instrument and its envisioned evolution will be at the forefront of underground instrumentation in fundamental research and they will foster research in key quantum technologies. Beyond the
development of this equipment, the results of the MIGA project will pave the way to enhancement technique for the low frequency sensitivity of future gravitational wave detectors.

\section*{Acknowledgments}

The MIGA Equipex is funded by ANR under the program \textit{Investissement d'Avenir} (ANR-11-EQPX-0028). We also  thank funding from the city of Paris (Emergence project HSENS-MWGRAV), D\'el\'egation G\'en\'erale pour l'Armement (DGA) and CNES.

\end{document}